\documentstyle[editedvolume,numreferences]{crckapb} 
\def\h100inv{$h_{\hbox{\sixrm 100}}^{\sixrm -1}$}         % Ho 100 inverse
\def\littleprime{\ifmmode{\scriptscriptstyle \prime }
    \else{\hbox{$\scriptscriptstyle \prime$ }}\fi}
\def\littlecirc{\ifmmode{\scriptscriptstyle \circ }
    \else{\hbox{$\scriptscriptstyle \circ $ }}\fi}
\def\littless{\ifmmode{\scriptscriptstyle s }
    \else{\hbox{$\scriptscriptstyle s $ }}\fi}
\def\arcss{\raise .9ex \hbox{\littless}}
\def\arcsec{\raise .9ex \hbox{\littleprime\hskip-3pt\littleprime}}
\def\arcmin{\raise .9ex \hbox{\littleprime}}
\def\degree{\raise .9ex \hbox{\littlecirc}}
\def\deg{\raise .9ex \hbox{\littlecirc}}
\def\arcsspoint{\hbox to 1pt{}\rlap{\arcss}.\hbox to 2pt{}}
\def\arcsecpoint{\hbox to 1pt{}\rlap{\arcsec}.\hbox to 2pt{}}
\def\arcminpoint{\hbox to 1pt{}\rlap{\arcmin}.\hbox to 2pt{}}

\begin{opening}
\title{THE CNOC2 FIELD GALAXY REDSHIFT SURVEY}

\author{H.K.C. Yee, M.J. Sawicki, R.G.~Carlberg, H.~Lin}
\institute{University of Toronto\\
      Dept. of Astronomy, Toronto ON M5S 3H8, Canada     }

\author{S.L.~Morris}
\institute{DAO/HIA,
      5071 Saanich Rd, Victoria BC V8X 4M6  Canada     }

\author{D.R.~Patton, G.D.~Wirth}
\institute{University of Victoria\\
      Dept. of Physics \& Astronomy, Victoria BC V8W 3P6, Canada}

\author{C.W.~Shepherd}
\institute{University of Toronto\\
      Dept. of Astronomy, Toronto ON M5S 3H8, Canada     }
\author{E. Ellingson}
\institute{University of Colorado\\
      CASA, CB 389, Boulder  CO 80309, USA }

\author{D. Schade, R. Marzke }
\institute{DAO/HIA
      5071 Saanich Rd, Victoria BC V8X 4M6  Canada     }

\end{opening}

\runningtitle{THE CNOC2 Redshift Survey}

\begin{document}

% The \begin{document} command comes after the \end{opening}
% command.

\section{Introduction}

Fundamental to our understanding of the universe is the evolution
of structures, from galaxies to clusters of galaxies to large-scale
sheets and filaments of galaxies and voids.
The investigation of the evolution of large-scale structure not
only provides us with the key test of theories of structure
formation, but also allows us to measure fundamental
cosmological parameters.
The CNOC2 (Canadian Network for Observational Cosmology) Field
Galaxy Redshift Survey is the first large redshift survey
of faint galaxies
carried out with the explicit goal of investigating the evolution
of large scale structure.
This survey also provides the largest redshift and photometric data
set currently available for the study of galaxy population and 
evolution at the moderate redshift range between 0.1 and 0.6.
In this paper we describe the scope and technique of the survey,
its status, and some preliminary results.

\section{Observational Strategy}

The CNOC2 survey is being conducted using the MOS arm of the MOS/SIS
spectrograph at the Canada-France-Hawaii Telescope.
The primary goal of the survey is to obtain a large enough sample of galaxies
with high quality spectroscopic and photometric data for the purpose of
studying the large scale structure at the redshift of $\sim0.35$.
We need to obtain a sample of galaxies comparable to the large
nearby redshift surveys (e.g., Geller \& Huchra 1989; Shectman et al.~1996).
This requires obtaining redshifts for the
order of 10$^4$ galaxies covering an area of sky
subtending well over 10 $h^{-1}$ Mpc with velocity measurements
accurate to better than 100 km s$^{-1}$.

To avoid being dominated by a small number of large structures, the
survey covers four widely separated regions, called {\it patches},
on the sky.
The patches are  distributed
so that two regions can be observed at any time of the year.
Each patch is a mosaic of 20 MOS fields, covering $\sim 1400$
sq arcmin.
The total area of the survey is $\sim 1.5$ sq degrees.
Each spectroscopically defined MOS field is $\sim$9\arcmin$\times$8\arcmin,
with about 15\arcsec overlapping area with the adjacent fields.
The overlapping of the fields minimizes the systematic uncertainties
in photometric calibrations from field to field.
The field layout, in a L shape with a central block, is shown in Figure 1.
The maximum dimensions of the patches span 
$\sim$ 80\arcmin NS, and 63\arcmin EW.
The patches are chosen to avoid bright ($<12$ mag) stars,
low-redshift (e.g., Abell) clusters, and other known low redshift
bright objects.
They have galactic latitudes between 45\deg and 60\deg, chosen to 
avoid  excessive Galactic extinction ($<0.05$ mag in $A_B$),
yet close enough to the galactic plane to provide
sufficient number of stars to overcome the 
star-galaxy classification problem for the MOS images which
 have variable point-spread-function across the field.

\begin{figure}
\includegraphics{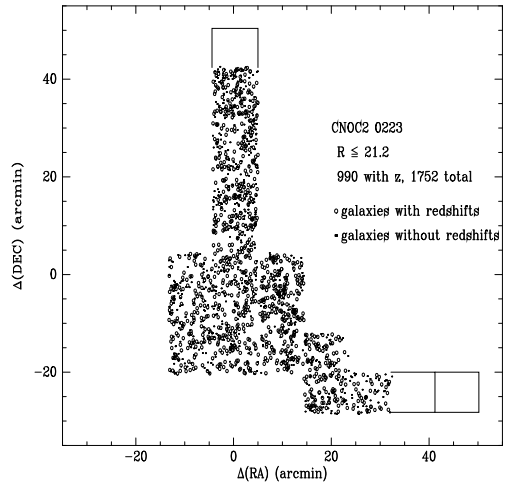}
\vspace{6.5cm}
\caption{Field layout in a patch. Currently, 17 of 20 fields in 0223+00
have been observed, with a total of 1310 redshifts.
Symbols denote galaxies in a $R\le21.2$ sample.}
\end{figure}

We utilize the observational technique developed for the CNOC1
Cluster Redshift Survey (Yee, Ellingson, \& Carlberg 1996; hereafter, YEC).
Here, we describe briefly the procedure and some improvements.
For the first 3 runs, two different CCDs with 15$\mu$/pixel were
used. However, for the remaining 4 runs, from which most of
the data were (or will be) obtained, a STIS CCD, which has
 higher quantum efficiency (QE) and  superior
blue response, but with a larger 21$\mu$ pixel size, was used.
Here, for brevity, we describe primarily the observational 
parameters for runs using the higher QE STIS CCD.

Images in $I$, $R$, Gunn $g$, $B$, and $U$ for each field are
obtained using the imaging mode of MOS with integration times of 6
to 15 min, obtaining
average 5$\sigma$ detection magnitudes for $R$ and $B$ of 
24.0 and 24.6, respectively.
The $B$ and $U$ images for all the fields
 are taken with the blue sensitive CCD.
The photometry is reduced and multicolor catalogs produced
in real time at the telescope. Using the
catalogs, multislit masks are designed
using a computer program which allows one to prioritize the sample in
various ways and optimize the number of slit placements.
The slits have a width of 1\arcsecpoint3 and a minimum length
of 11\arcsec.
The masks are cut using LAMA, a computer controlled laser cutting
machine, and are available for the spectroscopic observation within
as little as 3 hours after the direct imaging observation.
The primary spectroscopic sample is the union set of galaxies with
$R<21.5$ or $B<22.5$.
This allows us to construct unbiased $R$ and $B$ samples from the
redshift catalogs.

The spectroscopy observations are done using the B300 grism providing
a resolution of $\sim$15\AA.
A band limiting filter is used to shorten the spectrum so that
more slits can be placed on each mask, with typical numbers
ranging from 85 to 110.
The wavelength coverage of the filter is 4300\AA~to 6300\AA, allowing
for the unbiased coverage of important spectroscopic features within
the target redshift range of $0.10<z<0.55$ for the survey.

Two masks are used for each field.
Mask A covers primarily galaxies brighter than $R$=20.0, with
a total integration time of 40 min.
For mask B, which has an integration time of 80 min, the
highest priorities are assigned to 
galaxies with $R$ between 20 and 21.5.
Fainter objects are designed into the masks whenever possible.
The masks are designed so that
about 25\% of the galaxies observed spectroscopically are common to
both masks.
These redundant observations serve two purposes: first, for the fainter objects,
the summed spectra provide improved signal-to-noise ratio; second,
objects with redshifts determined independently from both masks provide
a rigorous empirical check for the velocity accuracy and the probability of
catastrophic redshift errors.
The two-mask strategy also allows us to compensate for the under-sampling
of close pairs that occurs when only one mask is used.

The total time required, using the more efficient CCD and including overhead, 
for the 5 colors and 2 spectroscopic masks per MOS field is about 3.5 hrs,
yielding typically 80 to 90 redshifts.
Thus, the complete survey of 80 fields
 requires about 27 clear nights, providing
about 6500 redshifts with a mean $z$ of 0.35.
At the intended survey spectroscopic limit of $R$=21.5, the 
cumulative sampling rate for the redshift sample is about 0.50 of
the photometric sample, while the differential completeness is about 0.25.

\begin{figure}
\includegraphics{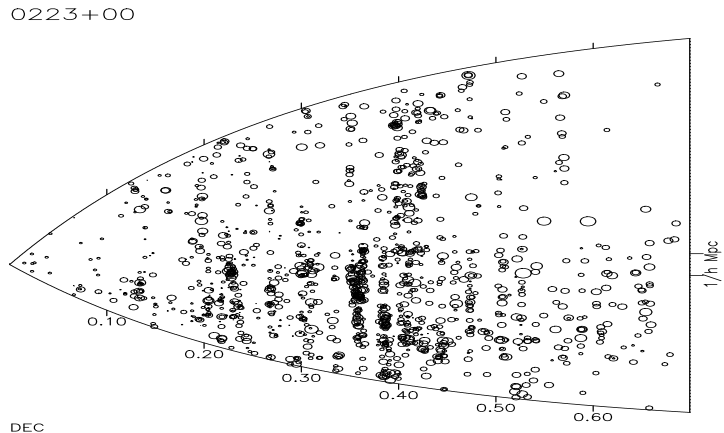}
\vspace{5.7cm}
\caption{Wedge Diagram for the patch 0223+00. 
 The tick marks on the vertical axis denote 1 $h^{-1}$ Mpc, while the
horizontal axis is marked by redshift. 
Note the sheet-like
structures spanning across the whole field in the Dec axis. 
The uneveness of the density
of points across the structures is due to an unequal numbers
of fields projected.}
\end{figure}

\section{Catalog Creation}

The data reduction procedure follows basically
 that outlined in YEC for the CNOC1 survey.
Photometry is performed using an improved version of PPP (Yee 1991).
Velocities are determined using cross-correlation.
With 5 photometric colors we are also able to verify the redshift
assignment.

Because the survey uses a sparse sampling strategy, accurate
weighting factors for each galaxy are essential for proper interpretation.
Although this has been discussed in detail in YEC where magnitude,
geometric, and color weights were presented, additional
effort has been put into better defining the selection function
 due to the limited redshift range over which the survey is sensitive.
Photometric redshifts will be used to estimate the survey completeness
as a function of redshift, and simulations of spectra will be used
to investigate how the redshift identification success rate depends
on variables such as spectral signal-to-noise ratio, galaxy type,
redshift, and surface brightness.
The objects from all the fields in each patch are 
merged into a single list after correcting for distortion
in the MOS image, creating a final catalog which includes
 photometry, redshift, and selection weights for
all objects brighter than $R=24$.

\begin{figure}
\includegraphics{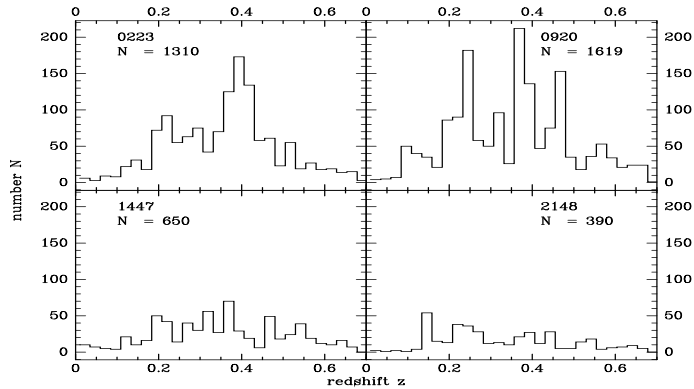}
\vspace{5cm}
\caption{Histograms of the $z$ distributions from the 3969
redshifts currently available in the 4 patches}
\end{figure}

\section{Survey Status}

The survey is presently (09/97) 80\% complete, and 
will be completed by the spring of 1998.
The current reduced sample contains $\sim$4000 redshifts, with an additional 
$\sim$1000 redshifts to be obtained from the data currently 
being reduced.
From the redundant observations the typical rms uncertainty of
the velocity determination is about 75 km s$^{-1}$ in the rest frame. 

Figure 1 illustrates the sky distribution of galaxies for the
redshift sample in the  17 fields currently available in the 0223+00 patch.
Figure 2 shows the wedge diagram of the 0223+00 patch.
Note the many sheet-like structures in redshift space spanning the
whole NS width of the patch of $\sim$10--15 $h^{-1}$ Mpc.
The redshift distributions of the current data set in the 4 patches
are shown as histograms in Figure 3.

Figure 4 shows some preliminary LFs obtained from a
subsample of 2768 galaxies with $R<21.2$.
The galaxies are divided into early, intermediate, and late types
based on $B-R$ colors.
Note the clear difference in the LFs for galaxies with different colors.
The CNOC2 sample is an order of magnitude larger than any
other redshift surveys in this redshift range (e.g., CFRS: Lilly
et al.~1995; Autofib: Ellis et al.~1996).
This large sample size,
in combination with extensive $UBgRI$ photometry (plus $K$ for some fields),
will permit precise measurements of the LF
and its evolution with redshift for different galaxy populations.

\begin{figure}
\includegraphics{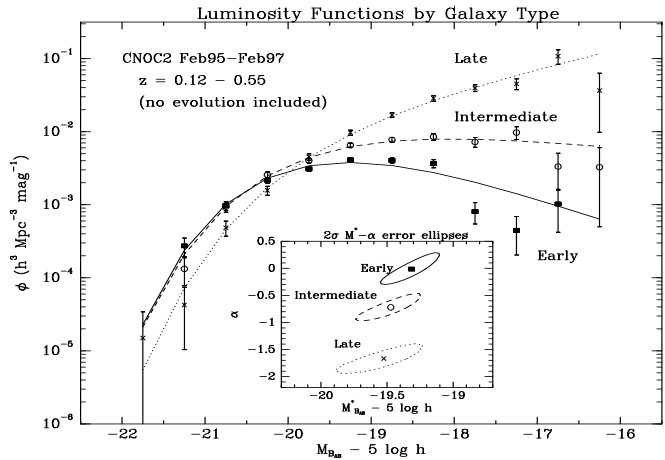}
\vspace{5.8cm}
\caption{Preliminary luminosity function of galaxies of different
population types as chosen by $B-R$ colors.}
\end{figure}

Currently, various projects are being carried out using this large
data base. We will be able to determine the galaxy correlation
function in redshift and morphological bins, estimate the evolution
of the pairwise velocity distribution, and measure the bias of galaxy
clustering relative to mass clustering.
The data are also being used to create a complete catalog of
groups and pairs of galaxies, from which we can study the dynamics
and evolution of galaxy groups and the redshift dependence of the merger
rate.
Line index measurements and principal component analysis of the spectra
will be carried out, allowing us to conduct a detailed investigation
of the star formation history of the galaxies.
In addition, we will also derive quantitative morphological parameters
of the galaxies, allowing us to examine the relationship between
morphology, environment, and galaxy evolution.
We will be able to obtain well-calibrated photometric redshifts for
about 20,000 galaxies down to $R\sim22.0$, which will further improve
our studies of galaxy clustering and populations, and also extend the
results to higher redshift.

\end{document}